\documentclass[prl,aps,twocolumn,10pt]{revtex4-1}
\usepackage{graphicx}
\usepackage{color}
\usepackage{amsmath}

\begin{document}

\title{Destination Choice Game: A Spatial Interaction Theory on Human Mobility}

\author{Xiao-Yong Yan$^{1,2}$}
\author{Tao Zhou$^{2}$}

\affiliation{
$^{1}$Institute of Transportation System Science and Engineering, Beijing Jiaotong University, Beijing 100044, China\\
$^{2}$Big Data Research Center, University of Electronic Science and Technology of China, Chengdu 611731, China
}

\begin{abstract}
With remarkable significance in migration prediction, global disease mitigation, urban plan{\color{black}ning} and many others, an arresting challenge is to predict human mobility fluxes between any two locations. A number of methods have been proposed against the above challenge, including the gravity model, the intervening opportunity model, the radiation model, the population-weighted opportunity model, and so on. Despite their theoretical elegance, all models ignored an intuitive and important ingredient in individual decision about where to go, that is, the possible congestion on the way and the possible crowding in the destination. Here we propose a microscopic mechanism underlying mobility decisions, named destination choice game (DCG), which takes into account the crowding effects resulted from spatial interactions among individuals. In comparison with the state-of-the-art models, the present one shows more accurate prediction on mobility fluxes across wide scales from intracity trips to intercity travels, and further to internal migrations. The well-known gravity model is proved to be the equilibrium solution of a degenerated DCG neglecting the crowding effects in the destinations.
\end{abstract}

\maketitle

Predicting human mobility fluxes between locations is a fundamental problem in transportation science and spatial economics \cite{OW11,RT03}. For more than a hundred years researchers have demonstrated the existence of gravity law in railway passenger movements \cite{O15,Z46}, highway car flow \cite{Z46,JWS08}, cargo shipping volume \cite{KKGB10}, commuters'  trips \cite{VBS06}, population migration \cite{TW95}, and so on. Therefore, the corresponding gravity model and its variants become the mostly widely used predictor for mobility fluxes and have found applications in many fields \cite{BBG17}, such as urban planning \cite{B08}, {\color{black} transportation science \cite{OW11,DLXD16},} infectious disease epidemiology \cite{FCFCCB06,LWD17} and migration prediction \cite{AS14}. However, the gravity model is just an analogy to the Newton's law, without any insights about the underlying mechanism leading to the observed mobility patterns. To capture the underlying mechanism of human mobility, some models accounting for individuals' decisions on destination choices were proposed, including the intervening opportunities (IO) model \cite{S40}, the  radiation model \cite{SGMB12} and the population-weighted opportunity (PWO) model \cite{YZFDW14,YWGL17}. { Some recently developed novel variants and extensions of the radiation and the gravity model \cite{SMN13,MSJB13,YHEG14,REWGT14,KLGQ15,BPTC16,VTN17,VTN18,CPGB18,LY19} can more accurately predict commuting, immigration or long distance travel patterns at different spatial scales. However, } all these models assume that individuals are independent of each other when selecting destinations, without any interactions.

In reality, individuals consider not only the destination attractiveness and the travelling cost, but also the crowding caused by the people who choose the same destination \cite{A94,CZ97,H18}, as well as the congestion brought by the people on the same way to the destination \cite{H18,HW97}. The crowding in the destination even happens in migration, because the more people move to a certain place, the competition among job seekers and the living expense become higher. For example, in China, the city with larger population are usually of higher house price. However, so far, to our knowledge, there is no mechanistic model about human mobility taking into account the crowding effects caused by spatial interactions among individuals.

In this paper, we propose a so-called destination choice game (DCG) to model individuals' decision-makings about where to go. In the utility function about destination choice, in addition to the travelling cost and the fixed destination attractiveness, we consider the costs resulted from the crowding effects in the destination and the congestion in the way. Extensive empirical studies from intracity trips to intercity travels, and further to internal migrations have demonstrated the advantages of DCG in accurately predicting human mobility fluxes between any two locations, in comparison with other well-known models including the gravity model, IO model{\color{black}, radiation model} and PWO model. We have further proved that the famous gravity model is equivalent to a degenerated DCG neglecting the crowding effects in the destination. Therefore, the higher accuracy of the prediction of DCG indicates the existence of the crowding effects on our decision-makings, which also provides a supportive evidence for the underlying hypothesis of the El Farol Bar problem \cite{A94} and the minority game \cite{CZ97}.

\section{Results}

\subsection{Model}

\begin{figure}
\center
\includegraphics[width=1.0\linewidth]{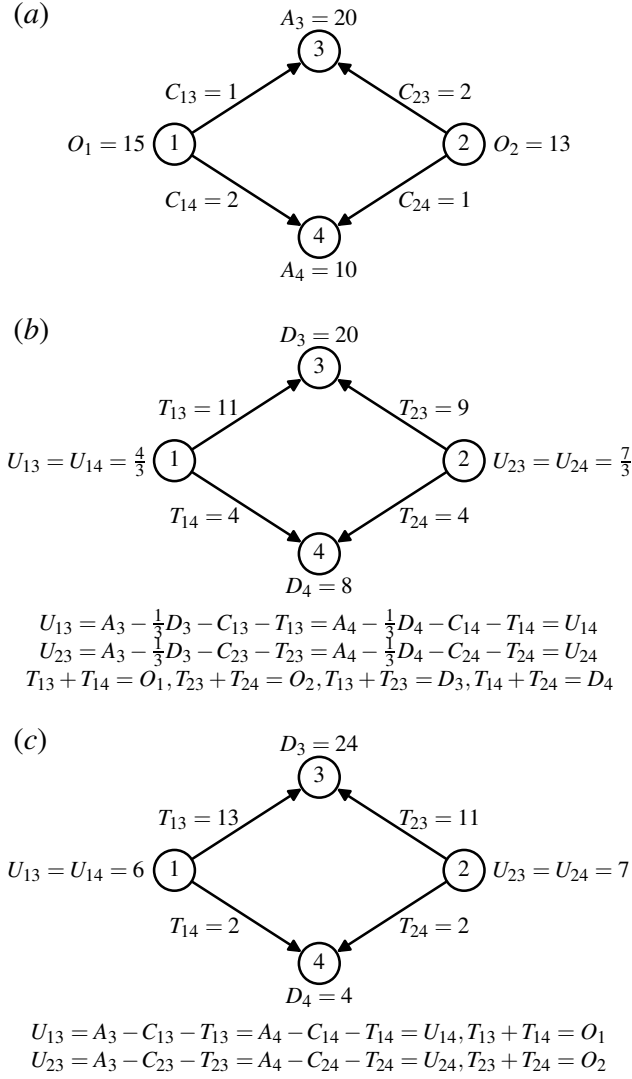}
\caption{ {\bf Illustration of a simple example of DCG.}
({\it a}) The game scene. The nodes 1 and 2 represent two starting locations while the nodes 3 and 4 are two destinations. $O_i$ is the number of individuals located in $i$, $A_j$ is the attractiveness of $j$, and $C_{ij}$ is the fixed travelling cost from $i$ to $j$. ({\it b}) An example game taking into account both the congestion effect on the way and the crowding effect in the destination, with a utility function $U_{ij}=A_j-\frac{1}{3}D_j-C_{ij}-T_{ij}$. ({\it c}) An example game that does not consider the crowding effect in the destination, with a utility function $U_{ij}=A_j-C_{ij}-T_{ij}$. For both ({\it a}) and ({\it b}), the equilibrium solutions are shown in the plots while the equations towards the solutions are listed below the plots.
}
\label{fig1}
\end{figure}

We introduce the details of the DCG model in the context of travel issues. { The number of individuals $T_{ij}$ travelling from the starting location $i$ to the destination $j$ is resulted from the cumulation of destination choices of all individuals at location $i$}. We model such decision-making process by a multiplayer game with spatial interactions, where each individual chooses one destination from all candidates to maximize his utility. Specifically speaking, the utility $U_{ij}$ of an arbitrary individual at location $i$ to choose location $j$ as destination consists of the following four parts. (i) The fixed payoff of the destination $h(A_j)$, where $h$ is intuitively assumed to be a monotonically increasing function of $j$'s attractiveness $A_j$ that is usually dependent on $j$'s population, GDP, environment, and so on \cite{L17}. (ii) The fixed travelling cost $C_{ij}$. (iii) The congestion effect $g(T_{ij})$ on the way, where { $T_{ij}$ is the target quantity} and $g$ is a monotonically non-decreasing function. (iv) The crowding effect $f(D_j)$ at the destination, where $f$ is a monotonically non-decreasing function and $D_j=\sum_i T_{ij}$ is the total number of individuals choosing $j$ as their destination. In a word, the utility function $U_{ij}$ reads
\begin{equation}
\label{eq1}
U_{ij}=h(A_j)-f(D_j)-C_{ij}-g(T_{ij}),
\end{equation}
where destination attractiveness $A_j$ and travelling cost $C_{ij}$ are input data, $T_{ij}$ is the model estimated flux from location $i$ to $j$ and destination attraction $D_j=\sum_i T_{ij}$.

In the above destination choice game (DCG), if every individual knows complete information, the equilibrium solution guarantees that all  $O_i$ individuals at the same starting location $i$ have exactly the same utility no matter which destinations to be chosen. Strictly speaking, the variable $T_{ij}$ has to be continuous to guarantee the existence of an equilibrium solution, which is a reasonable approximation when there are many individuals in each journey $i\rightarrow j$. Figure \ref{fig1}{\it a} illustrates a simple game scene. Considering a simple utility function $U_{ij}=A_j-\frac{1}{3}D_j-C_{ij}-T_{ij}$ that takes into account both the congestion effect on the way and the crowding effect in the destination, we can obtain the equilibrium solution based on the equilibrium condition ($U_{i3}=U_{i4}$) and the conservation law ($T_{i3}+T_{i4}=O_i$ and $T_{1j}+T_{2j}=D_j$). The solution is shown in Fig. \ref{fig1}{\it b}.

Generally speaking, we cannot obtain the analytical expression of the equilibrium solution, instead, we apply the method of successive averages \cite{BB06} (MSA, see {\bf Methods}) to iteratively approach the solution. {\color{black} Since the Weber-Fechner law \cite{T14} (see  {\bf Methods}) in behavioral economics is a good explanation of how humans perceive the change in a given stimulus, we select the logarithmic form determined by the Weber-Fechner law to} express the destination payoff function $h(A_j)$ as $\alpha\ln A_j$,
the destination crowding function $f(D_j)$ as $\gamma \ln D_{j}$ and
the route congestion function $g(T_{ij})$ as $\ln T_{ij}$.
{\color{black} On the other hand, s}ince travelling cost often follows an approximate logarithmic relationship with distance in multimodal transportation system \cite{Y13}, we use $\beta \ln d_{ij}$  instead of  $C_{ij}$,
 where $d_{ij}$ is the geometric distance between $i$ and $j$.
We then get a practical utility function
\begin{equation}
\label{eq2}
U_{ij} =\alpha\ln A_j -\beta \ln d_{ij}- \gamma \ln D_{j}-\ln T_{ij},
\end{equation}
where $\alpha$, $\beta$ and $\gamma$ are nonnegative parameters that can be fitted by real data {\color{black} (see  {\bf Methods})}, subject to the largest S{\o}rensen similarity index \cite{S48} (SSI, see  {\bf Methods}). $A_j$ is the location $j$'s attractiveness, which is approximated by the actual number of attracted individuals in the real data.

\begin{table*}
\centering
\caption{ {\bf Fundamental statistics of the data sets.} The second to fifth columns present the number of individuals, the number of recorded movements, the number of locations and how to estimate the geographical positions of these locations. For migration data, we do not know the precise number of individuals, but it should be close to the number of total records since people usually do not migrate frequently.}
\begin{tabular}
{p{5cm}p{3cm}p{3cm}p{3cm}p{3cm}}
\hline
data set & \#individuals & \#movements &  \#locations & positional proxy  \\
\hline
intracity trips in Abidjan & 154849 & 519710 &  381   &  base station \\
intercity travels in China & 1571056& 4976255 &   340 &  prefecture-level city \\
internal migrations in US & N/A & 2498464 &   51   &  state capital\\
\hline
\end{tabular}
\label{tb1}
\end{table*}

\subsection{Prediction}

We use three real data sets, including intracity trips in Abidjan, intercity travels in China and internal migrations in US, to test the predictive ability of the DCG model. The data set of intracity trips in Abidjan is extracted from the anonymous Call Detail Records (CDR) of phone calls and SMS exchanges between Orange Company's  customers in {\color{black}C\^{o}te d'Ivoire} \cite{BECC12}.  To protect customers' privacy, the customer identifications have been anonymized. The positions of corresponding base stations are used to approximate the positions of starting points and destinations. The data set of intercity travels in China \cite{YWGL17} is extracted from anonymous users' check-in records at Sina Weibo, a large-scale social network in China with functions similar to Twitter. Since here we focus on movements between cities, all the check-ins within a prefecture-level city are regarded as the same with a proxy position being the centre of the city. The data set of internal migrations in US is downloaded from https://www.irs.gov/statistics/soi-tax-stats-migration-data. This data set is based on year-to-year address changes reported on individual income tax returns and presents migration patterns at the state resolution for the entire US, namely for each pair of states $i$ and $j$ in US, we record the number of residents migrated from $i$ to $j$. The fundamental statistics are presented in Table \ref{tb1}. In all the above three data sets and other data sets presented in the {\bf Supplementary Information,  Table S1}, every location can be chosen as a destination.

\begin{figure*}
\center
\includegraphics[width=0.66\linewidth]{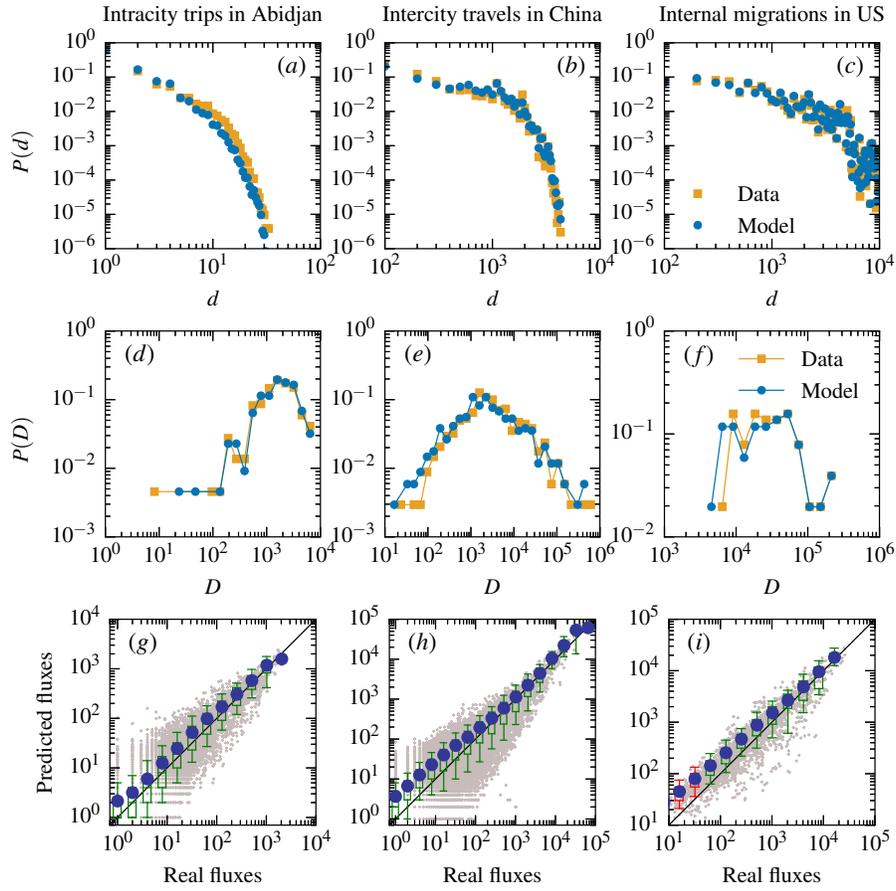}
\caption{{\bf Comparing the predictions of DCG model and the empirical data}.
({\it a}-{\it c}) Predicted and real distributions of travel distances $P(d)$.
({\it d}-{\it f}) Predicted and real distributions of locations's attracted travels $P(D)$.
({\it g}-{\it i}) Predicted and observed fluxes. The gray points are scatter plot for each pair of locations. The blue points represent the average number of predicted travels in different bins. The standard
boxplots represent the distribution of predicted travels in different bins.  A box is marked in green if the line $y=x$ lies between 10\% and 91\% in that bin and in red otherwise.
The data presented in (d-i) are binned using the logarithmic binning method.
}
\label{fig2}
\end{figure*}

\begin{figure*}
\center
\includegraphics[width=0.66\linewidth]{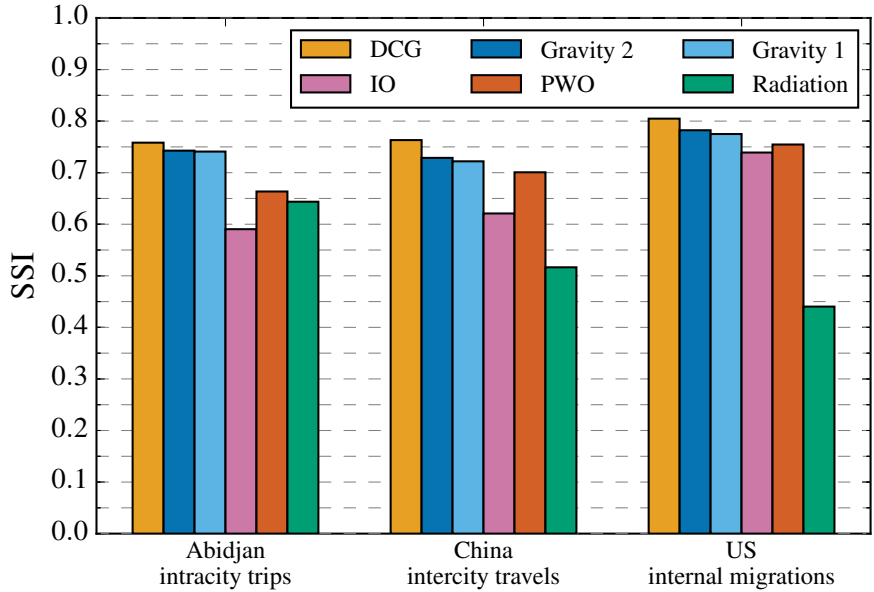}
\caption{ {\bf Comparing predicting accuracy of the DCG model and well-known benchmarks in terms of SSI.}
}
\label{fig3}
\end{figure*}

We use three different metrics to quantify the proximity of the DCG model to the real data. Firstly, we investigate the travel distance distribution, which is the most representative feature to capture human mobility behaviours~\cite{Y13,BHG06,GHB08}. As shown in Fig. \ref{fig2}{\it a}-{\it c}, the distributions of travel distances predicted by the DCG model are in good agreement with the real distributions. We next explore the probability $P(D)$ that a randomly selected location has eventually attracted $D$ travels (in the model, for any location $j$, $D_j$ is the total number of individuals choosing $j$ as their destination). $P(D)$ is a key quantity measuring the accuracy of origin-constrained mobility models, because origin-constrained models cannot ensure the agreement between predicted travels and real travels to a location~\cite{OW11}. Figure \ref{fig2}{\it d}-{\it f} demonstrate that the predicted and real $P(D)$ are almost statistically indistinguishable. Thirdly, we directly look at the mobility fluxes between all pairs of locations~\cite{SGMB12,YZFDW14,YWGL17}. As shown in Fig. \ref{fig2}{\it g}-{\it i}, the average fluxes predicted by the DCG model are in reasonable agreement with real observations.

We next compare the predicting accuracy on mobility fluxes of DCG with well-known models including the gravity models, the intervening opportunities model{\color{black}, the radiation model} and the population-weighted opportunities (PWO) model (see  {\bf Methods}). In terms of SSI, as shown in Fig. \ref{fig3}, DCG performs best. Specifically speaking, it is remarkably better than {\color{black} parameter-free models like the radiation model and }the PWO model and slightly better than the gravity model with two parameters.  {\bf Supplementary Information, Additional validation of the DCG model} shows extensive empirical comparisons between predicted and real statistics as well as accuracies of different methods for more data sets involving travels inside and  {\color{black} between} cities in Japan, UK, Belgium, US and Norway. Again,  {  in terms of SSI}, DCG outperforms other benchmarks in all cases. { Not only that, DCG also better predicts the travel distance distribution $P(d)$ and destination attraction distribution $P(D)$ in most cases (see Figs S5-S6 and tables S2-S3).}

\subsection{Derivation of the gravity model}
To further understand the advantage of the DCG model in comparison with the well-adopted gravity models, we give a close look at the key mechanism differentiated from all previous models, that is, the extra cost caused by the crowding effect, as inspired by the famous \emph{minority game} \cite{CZ97}. Accordingly, we test a simplified model without the term $f(D_j)$ in Eq. (\ref{eq1}). Figure \ref{fig1}{\it c} illustrates an example with a simple utility function $U_{ij}=A_j-C_{ij}-T_{ij}$ that only takes into account the congestion effect on the way. Similar to the case shown in Fig. \ref{fig1}{\it b}, the equilibrium solution can be obtained by the equilibrium condition and the conservation law. For a more general and complicated utility function (by removing the term related to the crowding effect in Eq. (\ref{eq2}))
\begin{equation}
\label{eq3}
U_{ij} =\alpha\ln A_j -\beta \ln d_{ij} - \ln T_{ij},
\end{equation}
based on the potential game theory \cite{MS96}, one can prove that the equilibrium solution is equivalent to the solution of the following optimization problem
\begin{equation}
\label{eq4}
\begin{aligned}
\max Z(\mathbf{x}) =& \sum\limits_{j} \int_0^{T_{ij}} (\alpha\ln A_j -\beta \ln d_{ij}-\ln x) \mathrm{d} x, \\
\mathrm{s.t.} \quad &\sum\limits_{j}T_{ij} = O_i, \ T_{ij} \geq 0.\\
\end{aligned}
\end{equation}
Since the objective function is strictly convex, the solution is existent and unique. Applying the Lagrange multiplier method, we can obtain the solution of Eq. (\ref{eq4}), which is exactly the same to the gravity model with two free parameters (i.e., Gravity 2, Eq. (\ref{eq11})), and if we set $\alpha=1$ in Eq. (\ref{eq3}), the solution degenerates to the gravity model with one free parameter (i.e., Gravity 1, Eq. (\ref{eq10})). The detailed derivation is shown in  {\bf Supplementary Information, Derivation of the gravity model using potential game theory}. The significance of such interesting finding is threefold. Firstly, it provides a theoretical bridge that connecting the DCG model and the gravity model, which are seemingly two unrelated theories. Indeed, it provides an alternative way to derive the gravity model. Secondly, comparing with the gravity models, the higher accuracy of the prediction from the DCG model suggests the existence of the crowding effect in our decision-making about where to go, which also provides a positive evidence for the validity of the critical hypothesis underlying the minority game. Thirdly, the improvement of accuracy from Gravity 2 to the DCG model can be treated as a measure for the crowding effect, which is, to our knowledge, the first quantitative measure for the crowding effect in human mobility.

\section{Discussion}

In summary, the theoretical advantages of DCG are twofold. First of all, it does not require any prerequisite from God's perspective, like the constraint on total costs in the maximum entropy approach \cite{W67,W10} and the deterministic utility theory \cite{N69}, or any oversubtle assumption, like the independent identical Gumbel distribution to generate the hypothetically unobserved utilities associated with travels in the random utility theory \cite{D75}. Instead, the two assumptions underlying DCG, namely (i) each individual chooses a destination to maximize his utility and (ii) congestion and crowding will {\color{black}decrease} utility, are very reasonable. Therefore, in comparison with the above-mentioned theories, DCG shows a more realistic explanation towards the gravity model by neglecting the crowding effect in destinations (see some other derivations to the gravity model in  {\bf Supplementary Information,  Other derivations of the gravity model}). Secondly, the present game theoretical framework is more universal and extendable. As the travelling costs and crowding effects are naturally included in the utility function, DCG is easy to be extended to deal with more complicated spatial interactions that depend on individuals' choices about not only destinations, but also departure time, travel modes, travel routes, and so on \cite{W52,V69,LSGHS16}. 
Not only that, the utility function of DCG can also be extended in predicting specific mobility behaviours. For example, when predicting the mobility fluxes in a multi-modal transportation system, the logarithmic (or linear logarithmic) function of distance is usually used to calculate the fixed travel cost between locations, while when predicting in a single-modal transportation system, the linear cost-distance function is usually used \cite{Y13}. For the destination payoff, destination crowding cost and route congestion cost in the utility function, {\color{black} although the DCG model has obtained better prediction accuracy by using the logarithmic functions inspired by the Weber-Fechner law, the realistic payoff and cost functions may be much more complicated. Therefore if we can mine real cost functions by some  machine learning algorithms from real data, the prediction accuracy could be further improved.}

In addition to theoretical advantages, DCG could better aid government officials in transportation intervention. For example, if the government would like to raise congestion charges in some areas (e.g., in Beijing, the parking fees in central urban areas are surprisingly high),  the parameter-free models like the radiation model and the PWO model cannot predict the quantitative impacts on travelling patterns since the population distribution is not changed, instead, the game theoretical framework could respond to the policy changes by rewriting its utility function. Another example is to forecast and regulate tourism demand \cite{S08}. In China, in the vacations of the National Day and the Spring Festival, many people stream in a few most popular tourist spots, leading to unimaginable crowding and great environmental pressure. Recently, Chinese government forecasts tourism demand before those golden holidays based on the booking information about air tickets, train tickets and entrance tickets, and then the visitors are effectively redistributed to more diverse tourist spots with remarkable decreases of visitors to the most noticed a few spots. Such phenomenon can be explained by the crowding effects in the destination choices, but none of other known models. In a word, DCG is more relevant to real practices and thus of potential to be enriched towards an assistance for decision making.

\section{Methods}

\subsection{Method of successive averages}
The method of successive averages (MSA)  is an iterative algorithms to solve various mathematical problems \cite{BB06}. For a general fixed point problem $\mathbf{x=F(x)}$, the $\mathbf{n}$th iteration in the MSA uses the current solution $\mathbf{x^{(n)}}$ to find a new solution $\mathbf{y^{(n)} = F(x^{(n)})}$. The next current solution is an average of these two solutions $\mathbf{x^{(n+1)}= (1-\lambda^{(n)}) x^{(n)} + \lambda^{(n)} y^{(n)}}$, where $0<\lambda^{\mathbf{(n)}}<1$ is a parameter. For the DCG model, the MSA contains the following steps:

{\bf Step 1}: Initialization. Set the iteration index $\mathbf{n}=1$. Calculate an initial solution { for the number of individuals travelling from $i$ to $j$}
\begin{equation}
\label{eq5}
T_{ij}^{\mathbf{(n)}} = O_i \frac{ A_j^{\alpha} d_{ij}^{-\beta}} {\sum_j A_j^{\alpha} d_{ij}^{-\beta}},
\end{equation}
where $O_{i}$ is {\color{black} an independent variable representing} the number of travellers starting from location $i$, $A_j$ is the attractiveness of location $j$ and $d_{ij}$ is the distance from $i$ to $j$ ($O_{i}$, $A_j$ and $d_{ij}$ are all initial input variables).

{\bf Step 2}:  Calculate a new solution { for the number of individuals travelling from $i$ to $j$}
\begin{equation}
\label{eq6}
F_{ij}^{\mathbf{(n)}} = O_i \frac{ A_j^{\alpha} d_{ij}^{-\beta} [D_{j}^\mathbf{(n)}]^{-\gamma}} {\sum_j A_j^{\alpha} d_{ij}^{-\beta}[D_{j}^\mathbf{(n)}]^{-\gamma}},
\end{equation}
where $D_j^{\mathbf{(n)}}=\sum_i T_{ij}^\mathbf{(n)}$  { is the total number of individuals choosing $j$ as their destination}.

{\bf Step 3}: Calculate the average solution
\begin{equation}
\label{eq7}
T_{ij}^{(\mathbf{n}+1)}= (1-\lambda^{\mathbf{(n)}}) T_{ij}^\mathbf{(n)} + \lambda^{\mathbf{(n)}} F_{ij}^\mathbf{(n)}.
\end{equation}
If $\vert T_{ij}^{(\mathbf{n}+1)}-T_{ij}^{\mathbf{(n)}} \vert < \varepsilon$ ($\varepsilon$ is a very small threshold, set as 0.01 in the work), the algorithm stops with current solution being the approximated solution; Otherwise, let $\mathbf{n}=\mathbf{n}+1$ and return to {\bf Step 2}.

For simplicity, we use a fixed parameter $\lambda^{\mathbf{(n)}}=\lambda = 0.5$.

\subsection{Weber-Fechner law}

Weber-Fechner Law (WFL) is a well-known law in behavioural psychology \cite{T14},  which represents the relationship between human perception and the magnitude of a physical stimulus. WFL assumes the differential change in perception $\mathrm{d} p$ to be directly proportional to the relative change $\mathrm{d}W/W$ of a physical stimulus with size $W$, namely  $\mathrm{d}p=\kappa \mathrm{d}W/ W$, where $\kappa$ is a constant. From this relation, one can derive a logarithmic function $p = \kappa  \mathrm{ln}(W / W_0)$, where $p$ equals the magnitude of perception, and the constant $W_0$ can be interpreted as stimulus threshold. This equation means the magnitude of perception is proportional to the logarithm of the magnitude of physical stimulus.  The WFL is widely used to determine the explicit quantitative utility function in behavioural economics \cite{T14} , and thus we adopt it in Eq. (\ref{eq2}).

\subsection{S{\o}rensen similarity index}

S{\o}rensen similarity index is a similarity measure between two samples \cite{S48}. Here we apply a modified version \cite{YZFDW14} of the index to measure whether real fluxes are correctly reproduced (on average) by theoretical models, defined as
\begin{equation}
\label{eq8}
\mathrm{SSI} = \frac{1}{N(N-1)}\sum^{N}_{i}{\sum^{N}_{j \neq i}{\frac{2 \min (T_{ij},T^{'}_{ij})}{T_{ij}+T^{'}_{ij}} }},
\end{equation}
where $T_{ij}$ is the predicted fluxes from location $i$ to $j$ and $T^{'}_{ij}$ is the empirical fluxes. Obviously, if each $T_{ij}$ is equal to $T^{'}_{ij}$ the index is 1, while if all $T_{ij}$ are far  from the real values, the index is close to 0.

{\color{black} 
\subsection{Parameter estimation}

We use grid search method  \cite{L80} to estimate the three parameters $\alpha$, $\beta$ and $\gamma$ of the DCG model. We first set the candidate value for each parameter from 0 to 10 at an interval of 0.01, and then exhaust  all the candidate parameter sets to calculate the SSI (see Eq. (\ref{eq8})) of the DCG model, and finally select the parameter set that maximizes SSI. The parameter estimation results are shown in {\bf Supplementary Information, Table S1}.
}

\subsection{Benchmark models}

We select two classical models, the gravity model and the intervening opportunities model, and {\color{black} two parameter-free models, the radiation model and} the population-weighted opportunities model, as the benchmark models for comparison with the DCG model.

(i) The gravity model is the earliest proposed and the most widely used spatial interaction model \cite{RT03}. The basic assumption is that the flow $T_{ij}$ between two locations $i$ and $j$ is proportional to the population $m_i$ and $m_j$ of the two locations and inversely proportional to the power function of the distance $d_{ij}$ between the two locations, as
 \begin{equation}
 \label{eq9}
T_{ij} = \alpha \frac{ m_{i}m_{j}}{ d_{ij}^{\beta}},
\end{equation}
where $\alpha$ and $\beta$ are parameters. To guarantee the predicted flow matrix $T$ satisfies $O_{i}=\sum_{j}T_{ij}$, we use two origin-constrained gravity models \cite{OW11}. The first one is called Gravity 1 as it has only one parameter, namely
\begin{equation}
\label{eq10}
T_{ij} = O_i \frac{ A_j d_{ij}^{-\beta}}{\sum_j A_j d_{ij}^{-\beta}},
\end{equation}
while the second one is named Gravity 2 for it has two parameters, as
\begin{equation}
\label{eq11}
T_{ij} = O_i \frac{ A_j^{\alpha} d_{ij}^{-\beta}}{\sum_j A_j^{\alpha} d_{ij}^{-\beta}}.
\end{equation}

(ii) The intervening opportunities (IO) model \cite{S40} argues that the destination choice is not directly related to distance but to the relative accessibility of opportunities to satisfy the traveller. The model's basic assumption is that for an arbitrary traveller departed from the origin $i$, there is a constant very small probability $\alpha/\beta$ that this traveller is satisfied with a single opportunity. Assume the number of opportunities at the $j$th location (ordered by its distance from $i$) is proportional to its population $m_{j}$, i. e. the number of opportunities is $\beta m_{j}$, and thus the probability that this traveller is attracted by the $j$th location is approximated $\alpha m_{j}$. Let $q_{i}^{(j)} = q_{i}^{(j-1)}(1-\alpha m_{j})$ be the probability that this traveller has not been satisfied by the first to the $j$th locations ($i$ itself can be treated as the 0th location), we can get the relationship $q_{i}^{(j)} = \mathrm{e}^{-\alpha S_{ij}}/(1-\mathrm{e}^{-\alpha M})$ between the probability $q_{i}^{(j)} $ and the total population $S_{ij}$ in the circle of radius $d_{ij}$ centred at location $i$,  where $M$ is the total population of all locations. Furthermore, we can get the expected fluxes from $i$ to $j$ is
\begin{equation}
\label{eq12}
T_{ij}=O_{i}(q_{i}^{(j-1)}-q_{i}^{(j)})=O_{i}\frac{\mathrm{e}^{-\alpha(S_{ij}-m_{j})}-\mathrm{e}^{-\alpha S_{ij}} }{ 1-\mathrm{e}^{-\alpha M} }.
\end{equation}

{\color{black}
(iii) The radiation model \cite{SGMB12} assumes that an individual at location $i$ will select the nearest location $j$ as destination, whose benefits (randomly selected from an arbitrary continuous probability distribution $p(z)$) are higher than the best offer available at the origin $i$.  The fluxes $T_{ij}$  predicted by the radiation model is
\begin{equation}
\label{eq13}
T_{ij}=O_{i}\frac{m_{i}m_{j}}{(S_{ij}-m_j)S_{ij}}.
\end{equation}

(iv)} The population-weighted opportunities (PWO) model \cite{YZFDW14}  assumes that the probability of travel from $i$ to $j$ is proportional to the attractiveness of destination $j$, inversely proportional to the population $S_{ji}$ in the circle centred at the destination with radius $d_{ij}$, minus a finite-size correction $1/M$. It results to the analytical solution as
\begin{equation}
\label{eq14}
T_{ij}=O_{i}\frac{m_{j}(\frac{1}{S_{ji}}-\frac{1}{M})}{\sum_{j }{m_{j}(\frac{1}{S_{ji}}-\frac{1}{M})}}.
\end{equation}
\\

\section{Acknowledgements} 
X.-Y.Y. was supported by NSFC under grant nos. 71822102, 71621001 and 71671015. T.Z. was supported by NSFC under grant no. 61433014.

\section{Contributions}

X.-Y.Y. and T.Z. designed the research; X.-Y.Y. and T.Z. performed the research; X.-Y.Y. analysed the empirical data; and T.Z. and X.-Y.Y. wrote the paper.

\section{Competing interests}

 The authors declare no competing interests.

\renewcommand\thesection{S\arabic{section}}
\renewcommand\theequation{S\arabic{equation}}
\renewcommand\thefigure{S\arabic{figure}}
\renewcommand\thetable{S\arabic{table}}
\makeatletter
 \renewcommand\@biblabel[1]{[S#1]}
 \renewcommand\@cite[1]{[S#1]}
\setcounter{figure}{0}
\setcounter{section}{0}
\setcounter{table}{0}
\makeatother

\newpage

\section{Supplementary Information}
\subsection{Additional validation of the DCG model}
 
We use two types of data, namely, intercity travels and intracity trips, to validate the DCG model. 
Description of these data sets is given below:

(1) Intercity travels. The data for intercity travels in Japan, U. K. and Belgium are extracted from the Gowalla check-in data set \cite{CML11-S} (https://snap.stanford.edu/data/loc-gowalla.html). Gowalla is a location-based social networking website on which users share their locations when checking in. The data set includes 6,442,890 check-ins of users over the period Feb. 2009 - Oct. 2010. For this data set, we define a user's travel as two consecutive check-ins in different cities. 

(2) Intracity trips. The records of intracity trips in New York and Los Angeles are extracted from the Foursquare check-in data set \cite{BZM12-S}, which contains 73,171 users. We define a user's trip as two consecutive check-ins at different locations (here, the locations are defined as the 2010 census blocks; see https://www.census.gov/geo/maps-data/maps/block/2010/). The total number of trips is 182,033.
The data for intracity trips in Oslo, Norway is extracted from the Gowalla check-in data set \cite{CML11-S}. Because of the absence of census blocks and traffic analysis zones in Oslo, we simply partition the city into 88 equal-area square zones, each of which is about 1 km $\times$ 1 km.  Each zone is one location in the city.

\begin{figure*}
\center
\includegraphics[width=0.7\linewidth]{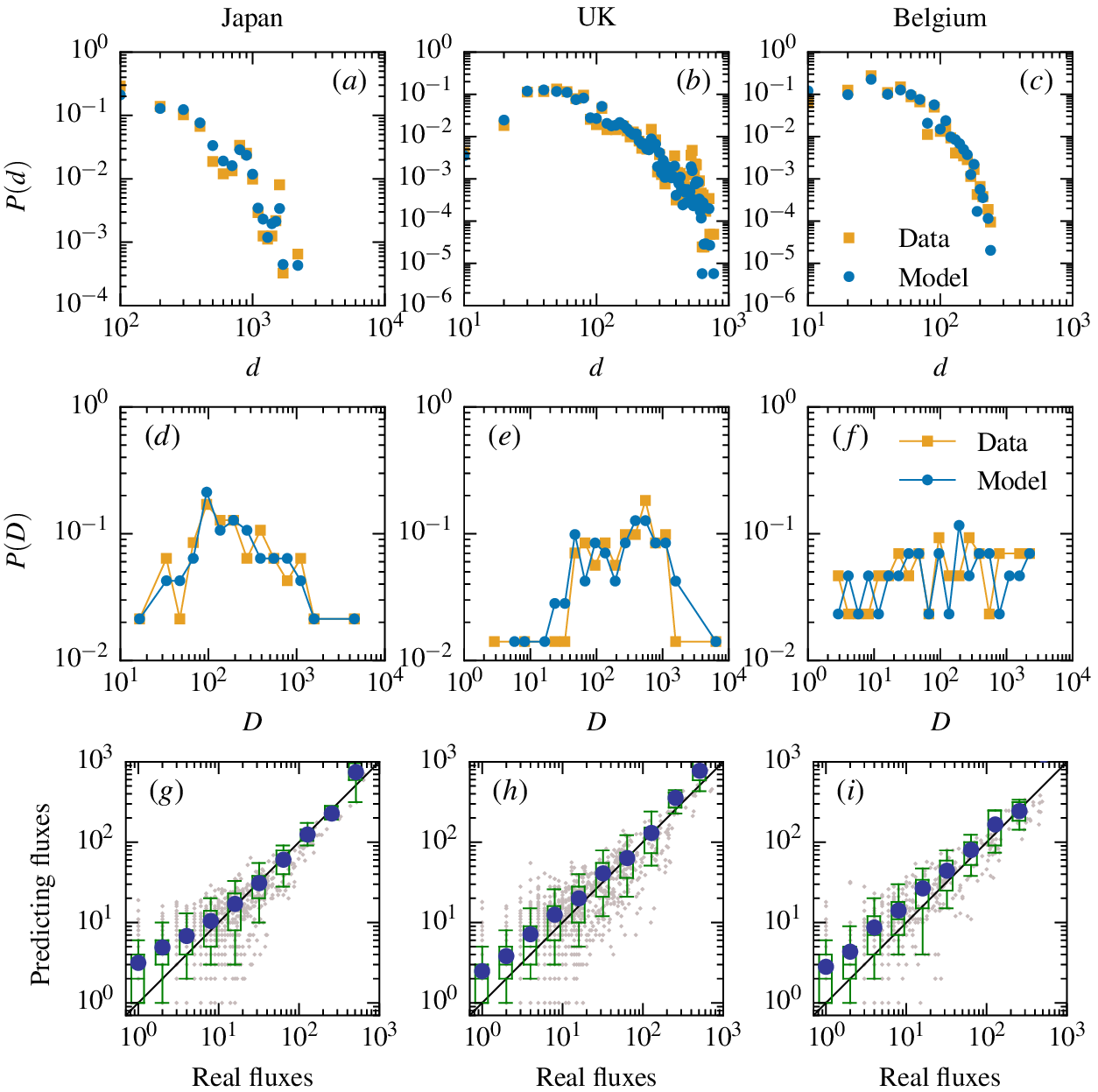}
\caption{{\bf Comparing the prediction of DCG model and the empirical data of intercity travels}.
({\it a}-{\it c}) Predicted and real distributions of travel distances $P(d)$.
({\it d}-{\it f}) Predicted and real distributions of locations's attracted travels $P(D)$.
({\it g}-{\it i}) Predicted and observed fluxes.
}
\label{figs1}
\end{figure*}

\begin{figure*}
\center
\includegraphics[width=0.7\linewidth]{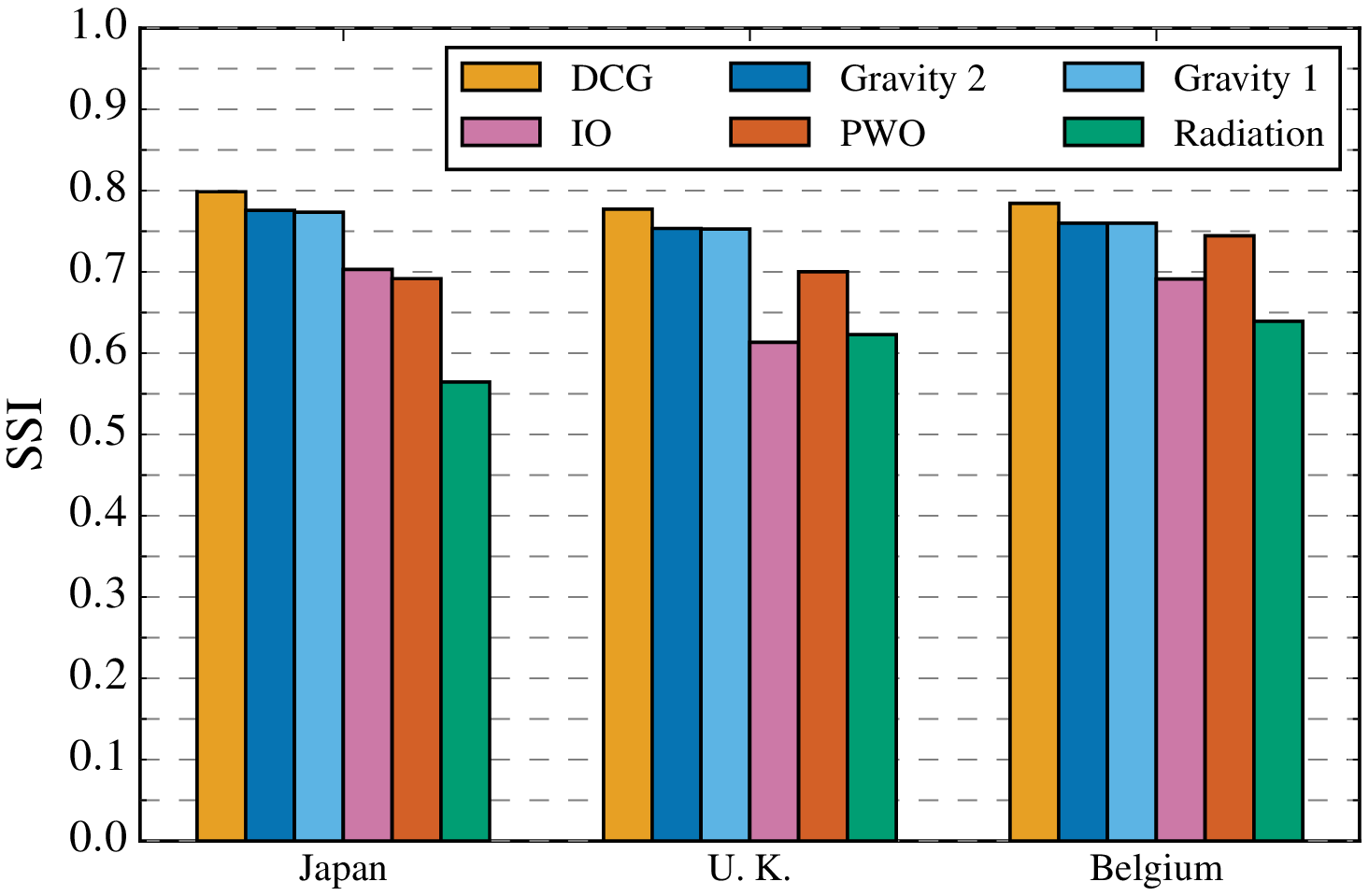}
\caption{ {\bf Comparing predicting accuracy of the DCG model and benchmark models for intercity travels.} 
}
\label{figs2}
\end{figure*}

\begin{figure*}
\center
\includegraphics[width=0.7\linewidth]{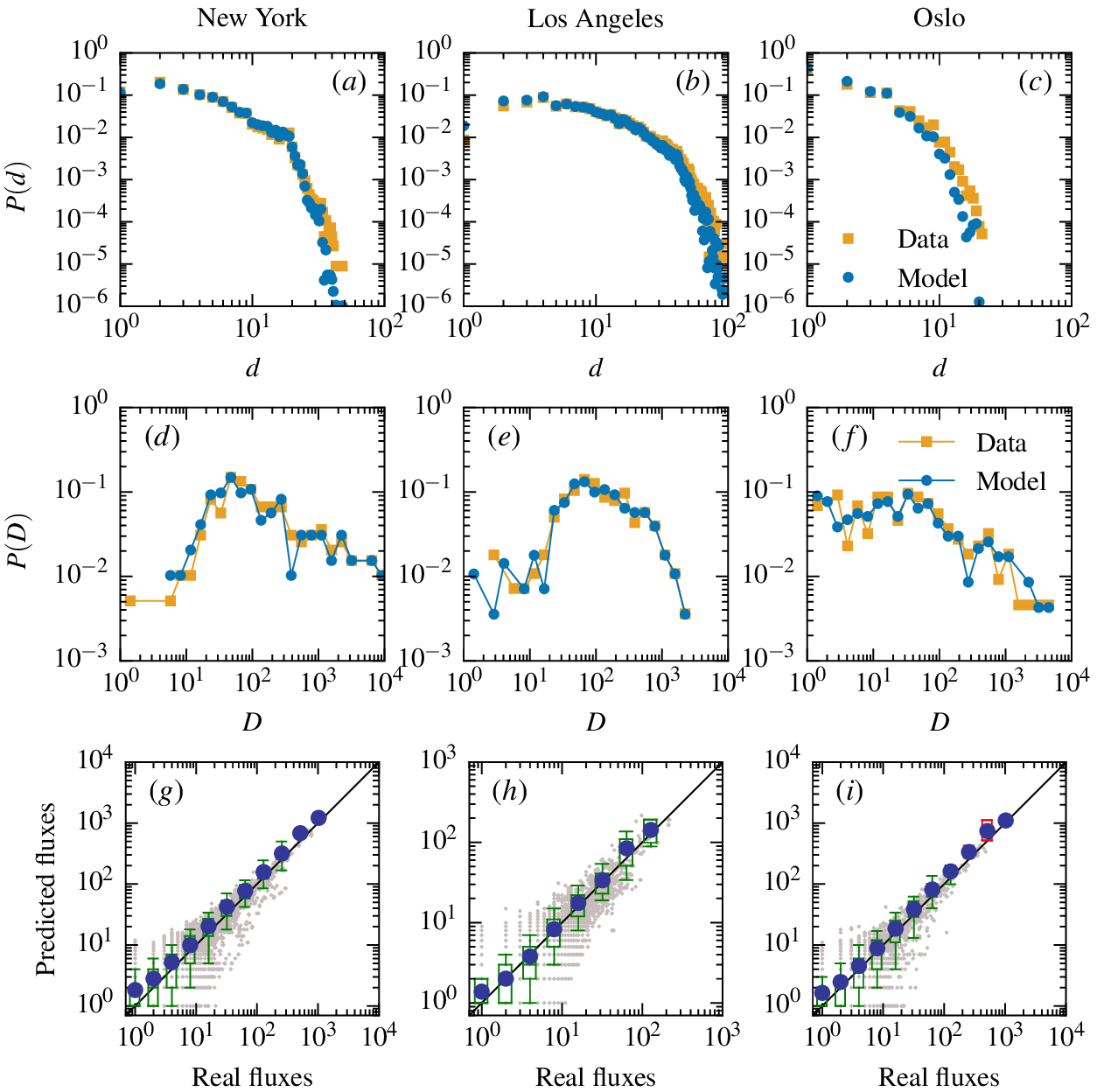}
\caption{{\bf Comparing the prediction of DCG model and the empirical data of intracity trips}.
({\it a}-{\it c}) Predicted and real distributions of travel distances $P(d)$.
({\it d}-{\it f}) Predicted and real distributions of locations's attracted travels $P(D)$.
({\it g}-{\it i}) Predicted and observed fluxes.
}
\label{figs3}
\end{figure*}

\begin{figure*}
\center
\includegraphics[width=0.7\linewidth]{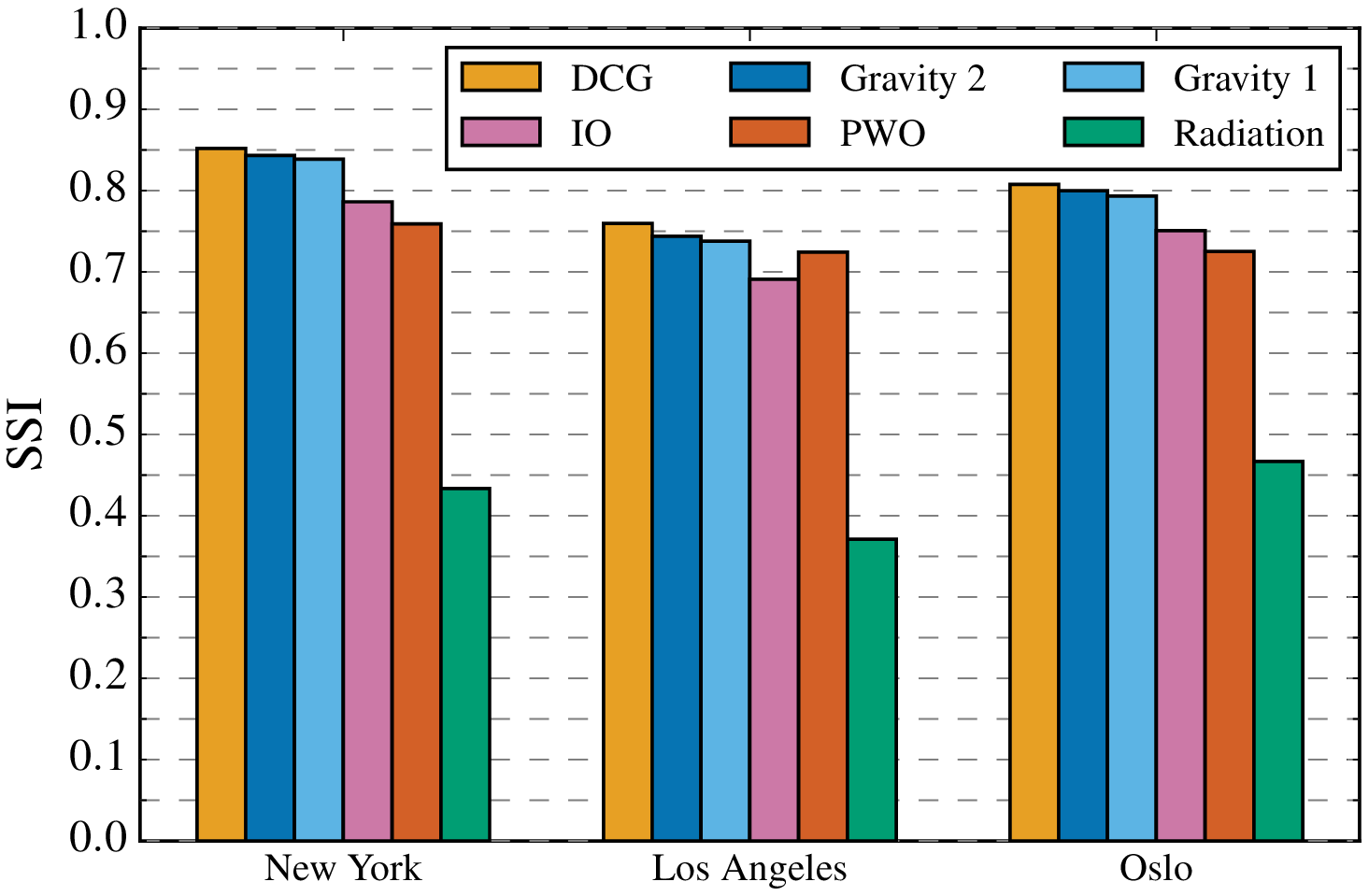}
\caption{ {\bf Comparing predicting accuracy of the DCG model and benchmark models for intracity trips.} 
}
\label{figs4}
\end{figure*}

The estimated model parameters  for these data sets are shown in table \ref{tbs1}, and the prediction results are shown in Figs \ref{figs1}-\ref{figs4}. Analogous to the results shown in the main text, DCG well predicts the real fluxes, with higher accuracy than other benchmarks subject to the SSI. 

\begin{figure*}
\center
\includegraphics[width=0.9\linewidth]{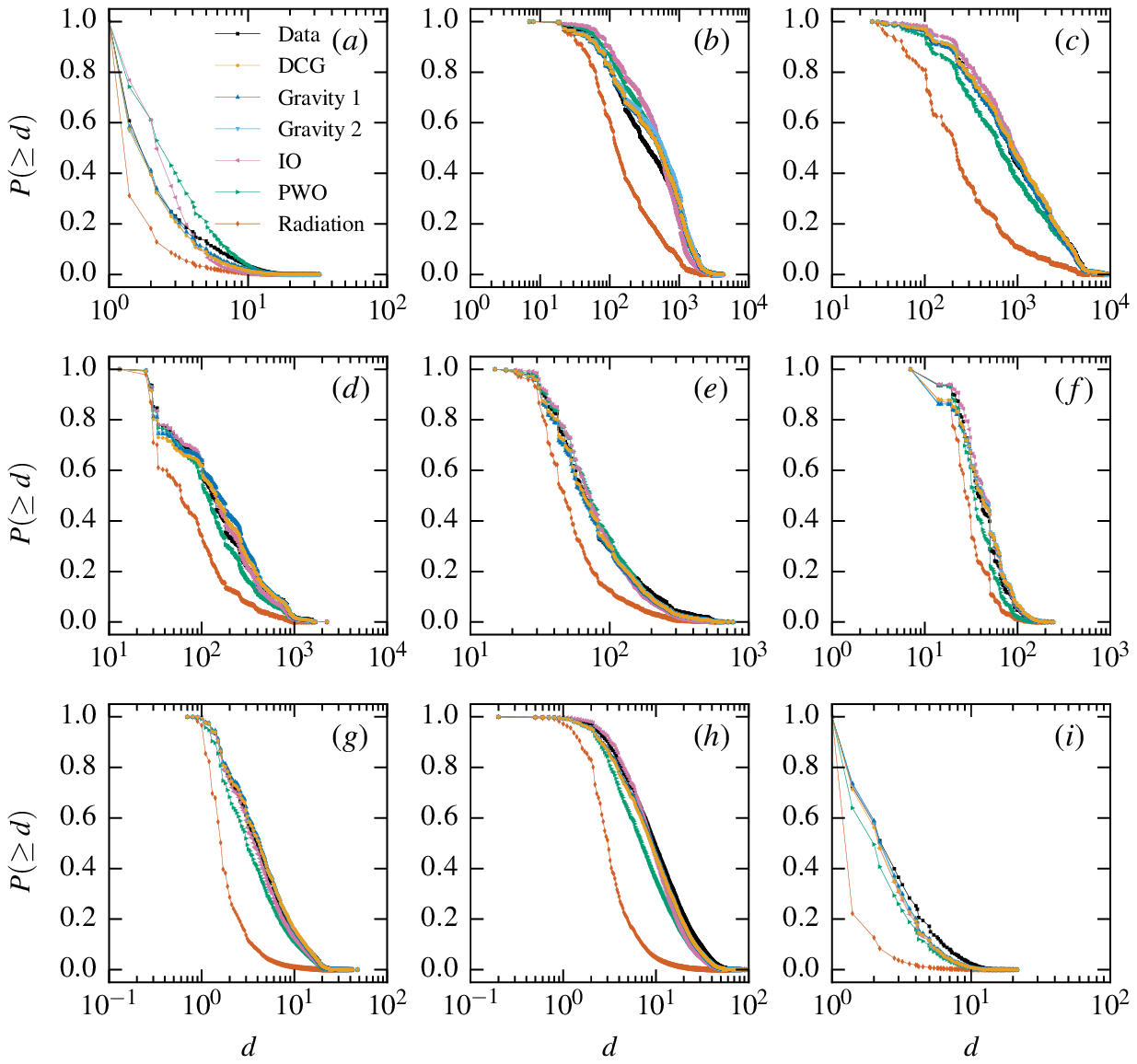}
\caption{  {\bf Comparing predicting accuracy of the DCG model and benchmark models in terms of travel distance distribution.} ({\it a}) Abidjan. ({\it b})  China. ({\it c}) US. ({\it d}) Japan. ({\it e}) UK. ({\it f}) Belgium. ({\it g}) New York. ({\it h}) Los Angeles. ({\it i}) Oslo.
}
\label{figs5}
\end{figure*}

\begin{figure*}
\center
\includegraphics[width=0.9\linewidth]{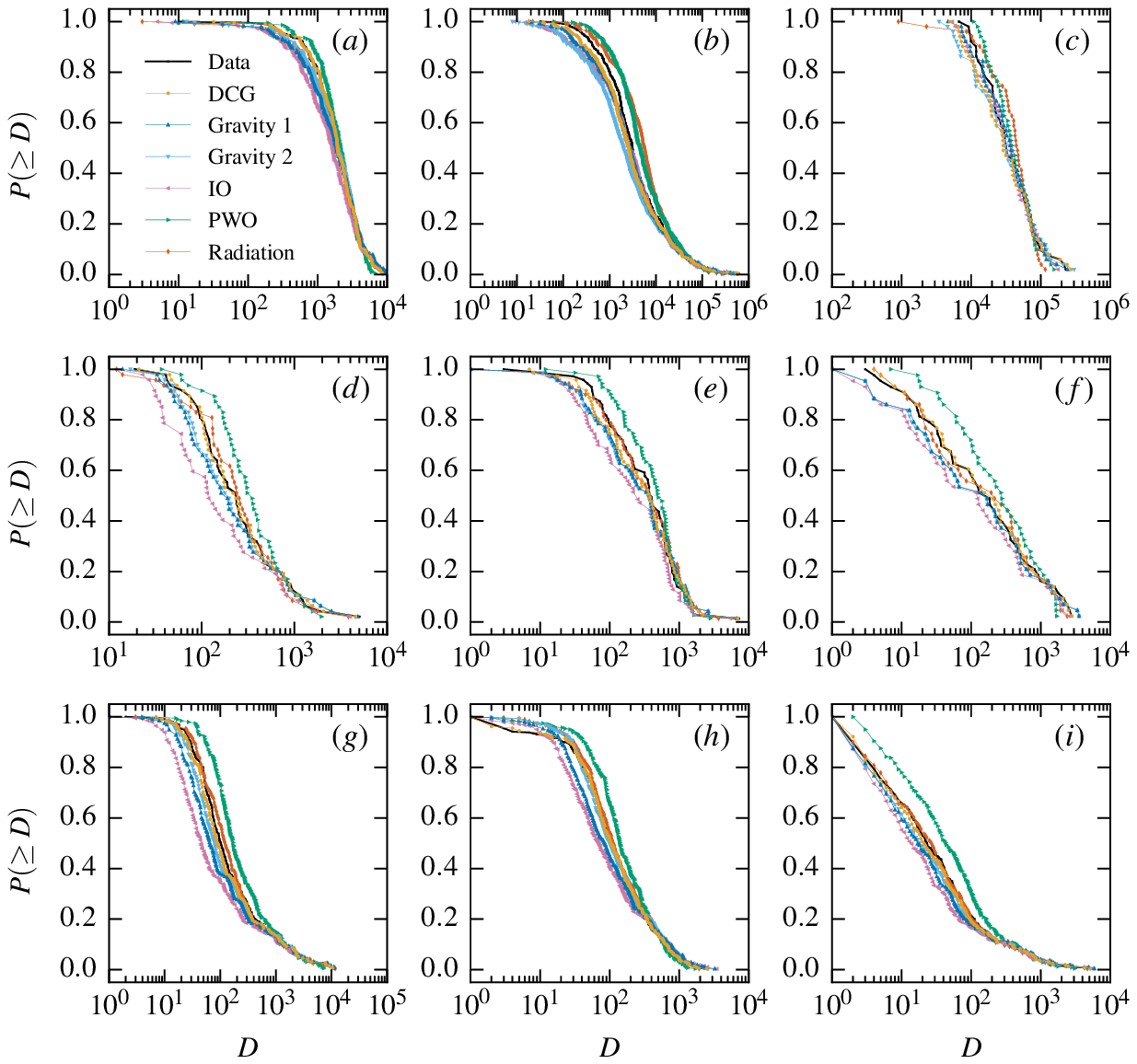}
\caption{  {\bf Comparing predicting accuracy of the DCG model and benchmark models in terms of destination attraction distribution.} ({\it a}) Abidjan. ({\it b})  China. ({\it c}) US. ({\it d}) Japan. ({\it e}) UK. ({\it f}) Belgium. ({\it g}) New York. ({\it h}) Los Angeles. ({\it i}) Oslo.
}
\label{figs6}
\end{figure*}

We further compare the predictions of the DCG model with other benchmark models in terms of travel distance distribution $P(d)$ and destination attraction distribution $P(D)$. The prediction results are shown in Figs \ref{figs5}-\ref{figs6}. In order to quantitatively compare the prediction accuracy of different models, we perform the two-sample  Kolmogorov-Smirnov (KS) test \cite{KS-S} on the model predicted and observed $P(d)$ and $P(D)$. The results are  shown in tables \ref{tbs2}-\ref{tbs3}, from which we can see that the KS statistics of the DCG model are generally smaller than or closer to that of the gravity model,  meaning that the DCG model has relatively high prediction accuracy. 

\subsection{Derivation of the gravity model using potential game theory}

Potential game theory originated from the congestion game presented by Rosenthal \cite{R73-S}. Monderer and Shapley defined exact potential games \cite{MS96-S} in which information concerning the Nash equilibrium can be incorporated in a potential function. They showed that every exact potential game is isomorphic to a congestion game. In the congestion game model, each player chooses a subset of resources. The benefit associated with each resource is a function of the number of players choosing it. The payoff to a player is the sum of the benefits associated with each resource in his strategy choice. A Nash equilibrium is a selection of strategies for all players such that no players can increase their payoffs by changing their strategies individually. Strategy profiles maximizing the potential function are the Nash equilibria \cite{VBVTF99-S}. 

From the introduction of the congestion game we can see that the degenerated destination choice game (DDCG) neglecting the crowding effect in the destination is a typical congestion game. Below we will give the process for finding the Nash equilibrium solution of the DDCG by maximizing the potential function of the congestion game.

A congestion game is a tuple $(N, R, (\Psi_k)_{k \in N}, (w_j)_{j \in R})$ \cite{VA06-S},  
where $N = \{1, \dots, n \}$ is a set of players 
(for DDCG, it is the set of $O_i$ travellers starting from origin $i$), 
$R = \{1, \dots, m\}$ is a set of resources 
(for DDCG, it is the set of destinations), 
$\Psi_k \subseteq 2^R$ is the strategy space of player $k$
(for DDCG, each player can only choose one destination in a strategy),
and $w_j$ is a benefit function associated with resource $j$
(for DDCG, it is the utility function, say $w_j = U_{ij}$).
Notice that benefit functions can achieve negative values, representing costs
of using resources \cite{VBVTF99-S}.
$S={S_1, \dots, S_n}$ is a state of the game in which 
player $k$ chooses strategy $S_k \in \Psi_k$.
For a state $S$,  the {\it congestion} $n_j(S) $ on resource $j$ is 
the number of players choosing $j$. 
For DDCG the number of travellers choosing destination $j$  is $T_{ij}(S)$.
The congestion game is an exact potential game \cite{MS96-S}, in which the potential function is defined as 
\begin{equation}
\label{eqpg1}
\phi(S) = \sum\limits_{j \in R} \sum\limits_{k=1}^{n_j(S)}w_j(k).
\end{equation}

For the DDCG  with utility function
$U_{ij} = \alpha \ln A_j - \beta \ln d_{ij}-\ln T_{ij}$, 
the potential function is
\begin{equation}
\label{eqpg2}
\phi(S) = \sum\limits_{j \in R} \sum\limits_{k=1}^{T_{ij}(S)}U_{ij}(k) = \sum\limits_{j \in R} \sum\limits_{k=1}^{T_{ij}(S)} (\alpha \ln A_j - \beta \ln d_{ij}-\ln k ),
\end{equation}
where $A_j$  is the attractiveness of location $j$,  $d_{ij}$ is the geometric distance between $i$ and $j$, and $\alpha$ and $\beta$ are nonnegative parameters.
To find the Nash equilibrium solution of DDCG, 
we treat $T_{ij}$ as a continuous variable. 
Then, Eq.  (\ref{eqpg2}) can be rewritten as
\begin{equation}
\label{eqpg3}
\phi(S) = \sum\limits_{j} \int_0^{T_{ij}(S)} (\alpha\ln A_j -\beta \ln d_{ij}-\ln x) \mathrm{d} x.
\end{equation}
 For the optimization problem in which $\max \phi(S)$ is subjected to  $\sum_j T_{ij} = O_i$, we can use the Lagrange multiplier method to obtain the solution. The Lagrangian expression is
  \begin{equation}
  \begin{aligned}
\label{eqpg4}
&\max L(T_{ij}, \lambda) = \\
&\sum\limits_{j} \int_0^{T_{ij}} (\alpha\ln A_j -\beta \ln d_{ij}-\ln x) \mathrm{d} x + \lambda (\sum\limits_j T_{ij} - O_i),
\end{aligned}
\end{equation}
where $\lambda$ is a Lagrange multiplier.  The partial derivative of the Lagrangian expression with respect to $T_{ij}$ is 
   \begin{equation}
\label{eqpg5}
\frac{\partial L}{\partial T_{ij}}= \alpha\ln A_j -\beta \ln d_{ij}-\ln T_{ij} + \lambda = 0,
\end{equation}
therefore
 \begin{equation}
\label{eqpg6}
T_{ij}  =\mathrm{e}^{ \lambda} A_j^{\alpha} d_{ij}^{-\beta}.
\end{equation}
Another partial derivative is 
\begin{equation}
\label{eqpg7}
\frac{\partial L}{\partial \lambda}= \sum\limits_j T_{ij} - O_i = 0.
\end{equation}
From Eq.  (\ref{eqpg6}) and Eq.  (\ref{eqpg7}) we can get
\begin{equation}
\label{eqpg8}
\sum\limits_j  \mathrm{e}^{ \lambda}  A_j^{\alpha} d_{ij}^{-\beta} - O_i = 0,
\end{equation}
or
\begin{equation}
\label{eqpg9}
\mathrm{e}^{ \lambda}  =  \frac{ O_i}{\sum_j A_j^{\alpha} d_{ij}^{-\beta}}
\end{equation}
By combining Eq.  (\ref{eqpg9}) and Eq.  (\ref{eqpg6}) we can derive
 \begin{equation}
\label{eqpg10}
T_{ij} = O_i \frac{ A_j^{\alpha} d_{ij}^{-\beta}}{\sum_j A_j^{\alpha} d_{ij}^{-\beta}},
\end{equation}
which happens to be an origin-constrained gravity model with two free parameters.
If we set $\alpha = 1$, the solution becomes
 \begin{equation}
\label{eqpg11}
T_{ij} = O_i \frac{ A_jd_{ij}^{-\beta}}{\sum_j A_jd_{ij}^{-\beta}},
\end{equation}
which is the standard origin-constrained gravity model \cite{OW11-S}.

Now back to the DCG model that considers 
both the congestion on the way and  the crowding in the destination.
Its utility function is
$U_{ij} =\alpha\ln A_j -\beta \ln d_{ij}- \gamma \ln D_{j}-\ln T_{ij}$.
If the crowding cost $\ln D_{j}$ is not affected by the fluxes $T_{ij}$,
the maximization of the potential function $\phi(S) = \sum_{j} \int_0^{T_{ij}} (\alpha\ln A_j -\beta \ln d_{ij} -\gamma \ln D_{j} -\ln x) \mathrm{d} x$  leads to the following result 
\begin{equation}
\label{eqpg12}
T_{ij} = O_i \frac{ A_j^{\alpha} d_{ij}^{-\beta} D_j^{-\gamma}}{\sum_j A_j^{\alpha} d_{ij}^{-\beta} D_j^{-\gamma}}.
\end{equation}
However, in fact, the destination attraction $D_j = \sum_i T_{ij}$ is dependent on the fluxes $T_{ij}$,  resulting in the essential difficulty in solving the Nash equilibrium of DCG . 
Therefore, we use the iterative algorithm MSA (see {\bf Material and Methods} in the main text) to numerically solve the DCG model.
In the MSA iteration, the function to calculate the iterative fluxes $F_{ij}^{(\mathbf{n})}$ is just the Eq. (\ref{eqpg12}).

If the destination attraction $D_j$ is fixed,
the DCG model's  potential function is 
$\phi(S) = \sum_{i} \sum_{j} \int_0^{T_{ij}} (\alpha\ln A_j -\beta \ln d_{ij} -\gamma \ln D_{j} -\ln x) \mathrm{d} x$.
For the optimization problem in which  $\max \phi(S)$ is subjected to  $\sum_j T_{ij} = O_i$ and $ \sum_i T_{ij} = D_j $,
the Lagrangian expression is
  \begin{equation}
\label{eqdoub1}
\begin{aligned}
&\max L(T_{ij})\\
 = & \sum\limits_{i} \sum\limits_{j} \int_0^{T_{ij}} (\alpha\ln A_j -\beta \ln d_{ij}- \gamma \ln D_{j}-\ln x) \mathrm{d} x  \\
& +\sum\limits_{i}\lambda_i (\sum\limits_j T_{ij} - O_i)  + \sum\limits_{j}\mu_j (\sum\limits_i T_{ij} - D_j) \\
=&\sum\limits_{j}\int_0^{D_{j}}(\alpha\ln A_j - \gamma \ln D_{j}) \mathrm{d} x \\
& -\sum\limits_{i} \sum\limits_{j} \int_0^{T_{ij}} (\beta \ln d_{ij}+\ln x) \mathrm{d} x \\
& +\sum\limits_{i}\lambda_i (\sum\limits_j T_{ij} - O_i)  + \sum\limits_{j}\mu_j (\sum\limits_i T_{ij} - D_j),
\end{aligned}
\end{equation}
where $\lambda_i$ and $\mu_j$ are Lagrange multipliers. The partial derivative of the Lagrangian expression with respect to $T_{ij}$ is 
   \begin{equation}
\label{eqdoub2}
\frac{\partial L}{\partial T_{ij}}= -\beta \ln d_{ij}-\ln T_{ij} + \lambda_i+ \mu_j = 0,
\end{equation}
therefore
 \begin{equation}
\label{eqdoub3}
T_{ij}  =\mathrm{e}^{ \lambda_i + \mu_j}  d_{ij}^{-\beta}.
\end{equation}
Since 
 \begin{equation}
\label{eqdoub4}
O_i = \sum\limits_j T_{ij} = \sum\limits_j \mathrm{e}^{ \lambda_i + \mu_j}  d_{ij}^{-\beta}
\end{equation}
and 
 \begin{equation}
\label{eqdoub5}
D_j = \sum\limits_i T_{ij} = \sum\limits_i \mathrm{e}^{ \lambda_i + \mu_j}  d_{ij}^{-\beta},
\end{equation}
we can get
 \begin{equation}
\label{eqdoub6}
\mathrm{e}^{ \lambda_i} = O_i /  \sum\limits_j \mathrm{e}^{ \mu_j}  d_{ij}^{-\beta}
\end{equation}
and
 \begin{equation}
\label{eqdoub7}
\mathrm{e}^{ \mu_j} = D_j /  \sum\limits_i \mathrm{e}^{ \lambda_i}  d_{ij}^{-\beta}.
\end{equation}
Let $a_i =\mathrm{e}^{ \lambda_i} / O_i$ and  $b_j =\mathrm{e}^{ \mu_j} / D_j $, 
Eq. (\ref{eqdoub3}) can be rewritten as 
 \begin{equation}
\label{eqdoub8}
T_{ij}  =a_i O_i b_j D_j d_{ij}^{-\beta},
\end{equation}
which is the standard doubly-constrained gravity model \cite{OW11-S}.
In the actual calculation, $a_i$ and $b_j$ are two sets of interdependent balancing factors, i.e. 
 $a_{i}=1/\sum_{j}b_{j}D_{j}d_{ij}^{-\beta}$ and $b_{j}=1/\sum_{i}a_{i}O_{i}d_{ij}^{-\beta}$.
This means that the calculation of one set requires the values of the other set: start with all $b_j = 1$, solve for $a_i=1/\sum_{j}b_{j}D_{j}d_{ij}^{-\beta}$ and
then use these values to re-estimate the $b_j=1/\sum_{i}a_{i}O_{i}d_{ij}^{-\beta}$; repeat until convergence of the two sets is achieved \cite{OW11-S}.

\subsection{Other derivations of the gravity model}

\subsubsection{Maximum entropy approach}

The earliest gravity model for spatial interaction was developed by analogy with Newton's  law of universal gravitation but lacked a rigorous theoretical base. Wilson proposed a maximum entropy approach to deriving the gravity model by maximizing the entropy of a trip distribution \cite{W67-S}

\begin{equation}
\label{eqme1}
\begin{aligned}
\max \ln \Omega =& \ \ln \frac{T!}{\prod_i \prod_j T_{ij}!} \\
\mathrm{s.t.} \quad &\sum\limits_{j}T_{ij} = O_i\\
& \sum\limits_{i}T_{ij} =D_i\\
& \sum\limits_{i}\sum\limits_{j}T_{ij}C_{ij} =C,\\
\end{aligned}
\end{equation}
where $\Omega$ is the number of distinct trip arrangements of individuals, $T$ is the total number of trips, $T_{ij}$ is the number of trips from location $i$ to location $j$, $O_i$ is the  total number of departures from $i$,  $D_j$ is the total number of arrivals at  $j$, $C_{ij}$ is the travelling cost from $i$ to $j$ and $C$ is the total travelling cost. 

According to the maximum entropy principle, the most likely trip distribution is the distribution with the largest number of microscopic states. Using the Lagrange multiplier method to solve Eq.  (\ref{eqme1}), we can get
\begin{equation}
\label{eqme2}
T_{ij} = a_{i} O_{i} b_{j} D_{j} \mathrm{e}^{-\lambda C_{ij}},
\end{equation}
where $a_{i}=1/\sum_{j}b_{j}D_{j}\mathrm{e}^{-\lambda C_{ij}}$ and $b_{j}=1/\sum_{i}a_{i}O_{i}\mathrm{e}^{-\lambda C_{ij}}$ are interdependent balancing factors.  Setting $\lambda C_{ij} = \beta \ln d_{ij} $, where $d_{ij}$ is the distance between $i$ and $j$, we can get a  doubly-constrained gravity model with power distance function
\begin{equation}
\label{eqme3}
T_{ij} = a_{i} O_{i} b_{j} D_{j} d_{ij}^{-\beta}.
\end{equation}

Wilson's maximum entropy derivation offers a theoretical base for the gravity model. However, the maximum entropy principle in statistical physics can only give the most likely macrostate (i.e., the most likely trip distribution matrix $\mathbf{T}$) but cannot describe the individuals' decision processes (i.e., the microscopic mechanism) in the system \cite{S78-S}. Meanwhile, the total cost in the maximum entropy method is not causally bounded by the theory itself, but determined externally \cite{HP79-S}. As the so-called total cost cannot be estimated in real world, the maximum entropy theory is less practical.

\subsubsection{Deterministic utility theory}

Some scientists described the micro decision-making process of  individual spatial interaction (destination choice) using the principle of utility maximization in economics \cite{S78-S}. Earlier studies used deterministic utility theory to derive the gravity model. The derivation is given in terms of trips made by individuals from a single origin to many destinations \cite{N69-S}. For an individual $k$ at origin $i$, assume that there are $\alpha m_j$ persons or things at each destination $j$   with which the individual at $i$ would like to interact per trip, where $m_j$ is the population at $j$ and $\alpha$ is a parameter.   Then, $k$'s utility of tripmaking from $i$ to all destinations is
\begin{equation}
\label{eqdu1}
 U_{i}^{(k)}  = \sum_j { U_{ij}^{(k)}} = \alpha \sum_j m_j f(T_{ij}^{(k)}),
\end{equation}
where $ U_{i}^{(k)}  $ is the total utility of individual $k$ at location $i$ of interactions with persons and
things at all destinations per unit time, $U_{ij}^{(k)}  $ is utility of interactions between  individual $k$ at location $i$ and persons or things at destination $j$ per unit time, $T_{ij}^{(k)}$ is the number of trips taken by individual  $k$ from  $i$ to  $j$ per unit time, and $f$ is a function.

An individual's number of trips is constrained by the total cost that the individual can pay,
\begin{equation}
\label{eqdu2}
r \sum_j { d_{ij} \  T_{ij}^{(k)}} \le \ C_{i}^{(k)},
\end{equation}
where $r$ is the cost per unit distance travelled and $C_i^{(k)}$ is the total amount of money individual $k$ located at $i$ is willing to spend on travels per unit time.

Setting $f(T_{ij}^{(k)})= \ln T_{ij}^{(k)}$ and using the Lagrange multiplier method to maximize Eq.  (\ref{eqdu1}) under constraint Eq.  (\ref{eqdu2}), we can derive
\begin{equation}
\label{eqdu3}
 T_{ij}^{(k)}  = \frac{ C_i^{(k)}}{r} \cdot \frac{m_j}{\sum\limits_j m_j} \cdot \frac{1}{d_{ij}}.
\end{equation}

The total number of trips taken by all individuals from $i$ to $j$ is obtained by summing the trips from  $i$ to  $j$ taken by all individuals at $i$:
\begin{equation}
\label{eqdu4}
 T_{ij}  =  \sum\limits_kT_{ij}^{(k)} = \frac{C_i}{r} \cdot \frac{m_j}{\sum\limits_j m_j} \cdot \frac{1}{d_{ij}},
\end{equation}
where $C_i$ is the total amount of money  that all individuals at origin $i$ are willing to spend
 on travels per unit time.

The main problem of this deterministic approach is that the total budget needs to be determined in advance. This is similar to the problem of Wilson's maximum entropy approach, which requires the prior constraint of the total cost. In addition, this method describes the individual's destination selection process over a continuous time period (i.e., the unit time). If the period is short enough and individuals can only complete one trip, then the individuals at a given origin will all select the same destination with the maximum utility, and there will be no dispersion of trips \cite{S78-S}.

\subsubsection{Random utility theory}

Domencich and McFadden applied the random utility theory to many transport-related discrete choice problems \cite{D75-S}, including trip destination choice. In this method, the random utility $U_{ij}$ of a destination $j$ for the individuals starting from origin $i$ is defined as
\begin{equation}
\label{eqru1}
 U_{ij}  = V_{ij}  +\varepsilon_{ij},
\end{equation}
where $V_{ij}$  is a nonstochastic element reflecting the observed attributes of $i$ and  $j$, and $\varepsilon _{ij}$ is a random variable describing an unobserved element containing attributes of the alternatives and characteristics of the individual that we are unable to measure.

The individual will choose the destination $j$ that maximized his utility, say
\begin{equation}
\label{eqru2}
U_{ij} > U_{ik} , \forall k \in J-\{j\},
\end{equation}
where $J$ is the set of all candidate destinations.

Since these utility values are stochastic, the choice probability of destination $j$ for any individual at $i$ is given by
\begin{equation}
\begin{aligned}
\label{eqru3}
P_{ij} & =  \mathrm{Prob}(U_{ij} > U_{ik}, \forall k \in J-\{j\})\\  & = \mathrm{Prob}(V_{ij} +\varepsilon_{ij}> V_{ik}+\varepsilon_{ik}, \forall k \in J-\{j\}) .
\end{aligned}
\end{equation}

If the random variables $\varepsilon _{ij}$  are independently and identically distributed Gumbel random variables, i.e.,
\begin{equation}
\label{eqru4}
F(\varepsilon _{ij})=\mathrm{e}^{-\mathrm{e}^{-\varepsilon _{ij}}},
\end{equation}
then, from Eq.  (\ref{eqru3}), we can get \cite{T09-S}
\begin{equation}
\label{eqru5-1}
\begin{aligned}
P_{ij}  &=\mathrm{Prob}(\varepsilon_{ik} <V_{ij} - V_{ik}+\varepsilon_{ij}, \forall k \in K-\{j\})\\
& = 
\mathrm{Prob}(\varepsilon_{ij}=x)  \cdot \prod_{k \neq j}  \mathrm{Prob}(\varepsilon_{ik}< V_{ij} -V_{ik} +x)\\
& = 
\int_{-\infty}^{\infty} \frac{\mathrm{d}  F(x)}{\mathrm{d} x}   F (V_{ij} -V_{ik} +x) \mathrm{d} x\\
& =\int_{-\infty}^{\infty}  \mathrm{e}^{- x} \mathrm{e}^{- \mathrm{e}^{-x}} \prod_{k \neq j} \mathrm{e}^{-\mathrm{e}^{- (V_{ij} - V_{ik}+x)}}\mathrm{d} x.
\end{aligned}
\end{equation}
Noting that $V_{ij} - V_{ij}=0$, so Eq.  (\ref{eqru5-1}) can be written as
\begin{equation}
\label{eqru5-2}
\begin{aligned}
P_{ij}  & =\int_{-\infty}^{\infty}  \mathrm{e}^{- x} \prod_{k } \mathrm{e}^{-\mathrm{e}^{- (V_{ij} - V_{ik}+x)}}\mathrm{d} x\\
& =\int_{-\infty}^{\infty} \mathrm{e}^{-\sum_k  \mathrm{e}^{- (V_{ij} - V_{ik}+x)}} \mathrm{e}^{- x} \mathrm{d} x\\
& =\int_{-\infty}^{\infty} \mathrm{e}^{-\mathrm{e}^{-x} \sum_k  \mathrm{e}^{- (V_{ij} - V_{ik})}} \mathrm{e}^{- x} \mathrm{d} x.
\end{aligned}
\end{equation}
Define $t=\mathrm{e}^{-x}$ such that $\mathrm{d}t=-\mathrm{e}^{-x}\mathrm{d}x$. When $x=\infty$, $t=\mathrm{e}^{-\infty}=0$ and when $x=-\infty$, $y=\mathrm{e}^{\infty}=\infty$. 
Therefore, Eq.  (\ref{eqru5-2}) can be written as
\begin{equation}
\label{eqru5-3}
\begin{aligned}
P_{ij}  & =\int_{\infty}^{0}   \mathrm{e}^{-t \sum_k  \mathrm{e}^{- (V_{ij} - V_{ik})}} \cdot (- \mathrm{d} t)\\
& =\int_{0}^{\infty}   \mathrm{e}^{-t \sum_k  \mathrm{e}^{- (V_{ij} - V_{ik})}} \mathrm{d} t\\
&=\frac{ \mathrm{e}^{-t \sum_k  \mathrm{e}^{- (V_{ij} - V_{ik})}} }{-\sum_k  \mathrm{e}^{- (V_{ij} - V_{ik})}} \Big\vert_0^{\infty}\\
&=\frac{ \mathrm{e}^{-\infty} -\mathrm{e}^0 }{-\sum_k  \mathrm{e}^{- (V_{ij} - V_{ik})}} \\
&=\frac{\mathrm{e}^{V_{ij}} }{\sum_k  \mathrm{e}^{V_{ik}}},
\end{aligned}
\end{equation}
which is the Logit model usually used in transport modal choice \cite{OW11-S}. If we set $V_{ij} = \ln m_j - \beta \ln d_{ij}$, we can get an origin-constrained gravity model
\begin{equation}
\label{eqru6}
T_{ij} = O_{i} P_{ij}  =O_{i}  \frac{\mathrm{e}^{V_{ij}}}{\sum\limits_j \mathrm{e}^{V_{ij}}} =O_{i}  \frac{m_j  d_{ij}^{-\beta}}{\sum\limits_j m_j  d_{ij}^{-\beta}}.
\end{equation}

Random utility theory accounts for the dispersion of trips from an origin and does not require a predetermined total budget.  Therefore, the gravity model based on random utility theory seems superior to other approaches based on deterministic utility theory or maximum entropy theory \cite{S78-S}. However, random utility theory asks for an oversubtle condition, namely the existence of an unobserved variable $\varepsilon_{ij}$ that has to obey the independent and identical Gumbel distribution.

\newpage

\begin{table*}\centering
\caption{{\bf Estimated parameters in parameterised spatial interaction models.} $\alpha$-DCG,   $\beta$-DCG and $\gamma$-DCG are the parameters in DCG model. $\alpha$-G2 and $\beta$-G2 are the parameters in Gravity 2 model. $\beta$-G1 is the parameter in Gravity 1 model and $\alpha$-IO is the parameter in IO model.}
\label{tbs1}
\begin{tabular}
{p{2.7cm}p{1.5cm}p{1.5cm}p{1.5cm}p{1.5cm}p{1.5cm}p{1.5cm}p{1.5cm}}
\hline
Data set & $\alpha$-DCG &  $\beta$-DCG & $\gamma$-DCG &  $\alpha$-G2 & $\beta$-G2 & $\beta$-G1 & $\alpha$-IO \\ 
\hline
Abidjan & 2.99& 2.53   & 2.28  & 0.88 & 2.49 & 2.43 & 3.04$\times 10^{-5}$\\
China  &  3.85 & 1.07 & 2.68 & 1.10 & 0.91& 0.96& 8.52$\times 10^{-7}$\\
US (migration) &  4.45 & 0.60  & 2.88 & 1.19 & 0.56 & 0.59& 7.73$\times 10^{-7}$\\
Japan   &  3.00   &  0.72    & 2.21   &  0.95  & 0.54   &  0.52 &  1.59$\times 10^{-4}$\\
UK   &  1.97   &  1.89    & 0.99   &  1.03  & 1.86   &  1.88  &  1.72$\times 10^{-4}$\\
Belgium  &  2.80  &  1.28   &  2.01  &  1.01  &  1.17  & 1.20   &  1.31$\times 10^{-4}$\\
New York   &  2.99  &  0.70   &  2.05  &  0.92  &  0.64  &  0.54  & 2.13$\times 10^{-5}$\\
Los Angeles   & 2.85   &  1.13   &  2.09  &  0.85  &  1.08  &  1.05  & 5.60$\times 10^{-5}$\\
Oslo &  1.86  &  1.03   & 1.01    &   0.90 &  0.87  &  0.76  & 5.87$\times 10^{-5}$\\
\hline
\end{tabular}
\end{table*}

\begin{table*}\centering

\caption{{\bf Two-sample  KS statistics of the travel distance distribution predicted by model and that observed from real data.} The bold number is the minimum value of KS statistic in a line.}
\label{tbs2}
\begin{tabular}{p{2.7cm}p{1.7cm}p{1.7cm}p{1.7cm}p{1.7cm}p{1.7cm}p{1.7cm}}
\hline
Data set & DCG &  Gravity 1  & Gravity 2 &  IO & PWO & Radiation\\ 
\hline
Abidjan               & 0.042   &  {\bf 0.032}  & 0.044 & 0.161  & 0.135& 0.296 \\
China                   & {\bf 0.077}    & 0.077 & 0.099 & 0.151  & 0.102& 0.286 \\
US (migration) & 0.031   & {\bf 0.023} &  0.031  & 0.052  & 0.090& 0.445 \\
Japan                  & 0.056    & 0.087 & 0.084 & {\bf 0.032}  & 0.068&  0.284\\
UK                         & {\bf 0.029}    & 0.048 & 0.044  & 0.056  & 0.061&  0.248\\
Belgium             & {\bf 0.064}    & 0.074 & 0.071 & 0.074  & 0.108&  0.282\\
New York           & {\bf 0.006}    & 0.006 & 0.007 & 0.012  & 0.023&  0.101\\
Los Angeles     & 0.005  & {\bf 0.003}  & 0.004  & 0.007  & 0.020 & 0.069\\
Oslo                    & 0.022  & 0.014   & {\bf 0.009} & 0.026  & 0.102&  0.500\\
\hline
\end{tabular}

\end{table*}

\begin{table*}\centering

\caption{{\bf Two-sample  KS statistics of the destination attraction distribution predicted by model and that observed from real data.} The bold number is the minimum value of KS statistic in a line.}
\label{tbs3}
\begin{tabular}{p{2.7cm}p{1.7cm}p{1.7cm}p{1.7cm}p{1.7cm}p{1.7cm}p{1.7cm}}
\hline
Data set & DCG &  Gravity 1  & Gravity 2 &  IO & PWO & Radiation\\ 
\hline
Abidjan                 &  {\bf 0.032 }  & 0.096  & 0.068 & 0.155  & 0.087  & 0.064 \\
China                    &  0.103   & {\bf 0.082}  & 0.171 & 0.103  & 0.147 &0.206\\
US (migration) & 0.137    & {\bf 0.098} &   0.157 & 0.118  & 0.137 &0.216\\
Japan                   &  {\bf 0.064}   & 0.191 & 0.149  & 0.277  & 0.277 &0.170\\
UK                          &  {\bf 0.085}    & 0.118 & 0.129 & 0.188  & 0.141 &0.102\\
Belgium               &  {\bf 0.070}  & 0.116  & 0.116  & 0.140  & 0.209 &0.093\\
New York            &  0.097    & 0.200 & 0.133 & 0.313  & 0.246 &{\bf 0.082}\\
Los Angeles       &  {\bf 0.024}   & 0.169   & 0.093& 0.220  & 0.164 &0.036\\
Oslo                      &  0.037   &0.100 &  0.074  & 0.140  & 0.154 &{\bf 0.026}\\
\hline
\end{tabular}

\end{table*}

\end{document}